\documentclass[
reprint,
superscriptaddress,
amsmath,amssymb,
aps,
prb,
]{revtex4-2}
\usepackage[english]{babel}
\usepackage[dvipsnames]{xcolor}
\usepackage{graphicx}%
\usepackage[caption=false]{subfig}
\usepackage{dcolumn}%
\usepackage{bm}%
\usepackage{hyperref}%
\hypersetup{
  colorlinks,
  citecolor=blue,
  linkcolor=blue,
  urlcolor=blue}

\newcommand{\beginsupplement}{
        \clearpage
        \setcounter{table}{0}
        \renewcommand{\thetable}{S\arabic{table}}
        \setcounter{section}{0}
        \renewcommand{\thesection}{S\arabic{section}}
        \setcounter{figure}{0}
        \renewcommand{\thefigure}{S\arabic{figure}}
        \setcounter{equation}{0}
        \renewcommand{\theequation}{S\arabic{equation}}
        \setcounter{page}{1} %
        \renewcommand{\thepage}{S\arabic{page}} %
        \onecolumngrid
}

\usepackage{float}

\newcommand{\yy}[1]{\textcolor{ForestGreen}{{\bf Yifan: #1 }}}

\begin{document}
\title{Ultrafast electron dynamics of electron-irradiated graphene}
\author{Yifan Yao}
\affiliation{Department of Materials Science and Engineering, University of Illinois, Urbana-Champaign, Urbana, IL 61801, USA}
\author{Andr\'e Schleife}
\email{schleife@illinois.edu}
\affiliation{Department of Materials Science and Engineering, University of Illinois, Urbana-Champaign, Urbana, IL 61801, USA}
\affiliation{Materials Research Laboratory, University of Illinois at Urbana-Champaign, Urbana, IL 61801, USA}
\affiliation{National Center for Supercomputing Applications, University of Illinois at Urbana-Champaign, Urbana, IL 61801, USA}

\begin{abstract}
Electron irradiation is essential for materials characterization and modification, though the fundamental interactions between incident electrons and host materials remain under investigation. Here, we employ first-principles simulations to study electron dynamics under external electron irradiation. We quantify differences in key observables, including kinetic energy loss, secondary electron emission, and backscattered electrons, between classical and quantum mechanical descriptions of the incident electron. 
Around 400 eV incident energy, we identify significant differences in backscattered electron yields between classical point-charge and quantum wave-packet descriptions, whereas the quantum-mechanical effects diminish at incident energies above 600 eV. 
These differences highlight the critical importance of quantum effects in electron irradiation phenomena that occur in a specific energy range of the incident electron. Our results provide clear guidance for selecting appropriate incident, electron descriptions based on kinetic-energy regimes, identify a targeted experimental window for isolating ``quantum-only" backscattering, and enable the rational design of 2D materials for nanofabrication and high-resolution electron-beam technologies.
\end{abstract}

\maketitle

\section{Introduction}

Electron irradiation induces crucial phenomena for material characterization and modification.
For materials characterization, electron microscopy is routinely employed for imaging with high spatial resolution \cite{ciston_surface_2015, meyer_imaging_2008, huang_grains_2011, huang_imaging_2013, wu_dissolution_2017, lu_fast_2016} down to sub Angstrom \cite{nguyen_achieving_2024}. 
Scanning electron microscopy relies on secondary electron emission to image surface morphology with nanoscale precision using electron energies ranging from a few hundred eV to tens of keV~\cite{goldstein_scanning_2003}. 
Beyond microscopy, electron beam lithography utilizes focused electron beams to fabricate nanoscale patterns~\cite{jesse_directing_2016, dyck_atom-by-atom_2019, utke_gas-assisted_2008, van_dorp_critical_2008}. 
Secondary electron emission is critical for ion sputtering coefficients, as the electron dynamics and ionization probabilities during irradiation fundamentally govern the yield and energy distribution of sputtered material~\cite{retsuo_kawakami_simultaneous_1999,schwarz_application_1990,schlueter_absence_2020}.
The versatility of tailoring electron kinetic energy for specific applications, such as surface imaging or nanofabrication, makes electron irradiation indispensable in modern materials manufacturing and characterization.

The success of electron microscopy and lithography has driven the development of theoretical simulations to explore electron-material interactions across multiple length and time scales.
Mesoscale modeling of these processes relies on knowledge from simulations of the underlying electronic structure and its time dependence~\cite{wang_modification_2012,rodgers_predicting_2016,zakirov_predictive_2020,yasuda_molecular_2007}.
However, such multiscale modeling is hindered by the complicated many-body and nonequilibrium nature of the underlying electron scattering phenomena.
To improve the accuracy of simulating electron scattering, efforts have been made to incorporate exchange-correlation effects beyond the adiabatic approximation~\cite{lacombe_electron_2018, suzuki_exact_2017, suzuki_machine_2020, van_faassen_time-dependent_2007}.
Additionally, first-principles simulations have been used to predict phenomena such as low-energy electron diffraction~\cite{yan_time-domain_2011, tsubonoya_time-dependent_2014} and secondary electron emission~\cite{ueda_quantum_2016, ueda_time-dependent_2018, miyauchi_electron_2017}, which are vital for materials imaging.
Furthermore, theoretical simulations aimed at optimizing nanoscale energy deposition~\cite{yao_first-principles_2021, gao_real-time_2025,tonzani_low-energy_2006} enhance the effectiveness of focused electron beam techniques.

For such studies, it is crucial to understand whether the resulting observables can be sufficiently captured by a classical description of the projectile electron.
Calculating observables of a classical point charge has been widely studied:
For example, the energy deposition of a classical point charge is described by stopping power, which has been studied for decades~\cite{schleife_accurate_2015, correa_calculating_2018, correa_nonadiabatic_2012, nordlund_molecular_1995, ullah_core_2018,sand_heavy_2019}.
However, computing electronic stopping for quantum electrons is more complicated both for numerical as well as fundamental reasons.
One numerical complication includes that the total energy of the electron subsystem, composed of target and projectile, is conserved.
Besides, for a quantum mechanical projectile, the kinetic energy cutoff of the (plane-wave) simulation has to be higher than the kinetic energy of the incident electron, which further requires a reduced time-step size in simulating the real-time dynamics.  

In addition, a comprehensive picture of how incident electrons deposit energy into a target and how this governs the subsequent electron emission remains incomplete due to the quantum nature of the incident electrons itself:
their de Broglie wavelength is comparable to atomic spacings, suggesting that quantum phenomena are critical.
This quantum description fundamentally differs from classical treatments and cannot be recovered merely through sampling of classical ensembles, which we will discuss using our results.
In particular, we uncover a purely quantum mechanical contribution to backscattering, opening up an experimental route to generate backscattered electrons that have no classical analog, which can be used to probe interesting quantum mechanical phenomena.
For these reasons, fundamental first-principles simulations that explicitly and quantitatively compare classical point‐charge versus quantum wave‐packet descriptions are required.

In this work, we comprehensively analyze femtosecond electron dynamics in graphene under electron irradiation using first-principles simulations, from the kinetic energy loss of the electron projectile to the subsequent electron emission of the host electrons. 
The extensive interest in graphene enables us to compare our results with prior studies on graphene under various projectiles~\cite{ueda_quantum_2016, yao_nonequilibrium_2024, kononov_anomalous_2021, vazquez_electron_2021, kononov_first-principles_2022, he_electronic_2021, zhang_ab_2012, ueda_time-dependent_2018, miyauchi_electron_2017,guichard_fast_2025}.
We employ real-time time-dependent density functional theory (rt-TDDFT), which offers a powerful framework for describing kinetic energy loss from the incident electron after impacting graphene.
This kinetic energy loss is analogous to computing electronic stopping power in bulk materials~\cite{schleife_accurate_2015,yost_examining_2017,lee_hot-electron-mediated_2019,lee_multiscale_2020}, and electron emission induced by ion irradiation~\cite{kononov_pre-equilibrium_2020,kononov_anomalous_2021, yao_nonequilibrium_2024, zhang_ab_2012}.
In this work, we seek to establish a detailed connection between kinetic energy loss, electron emission yield, and the kinetic energy distribution of the backscattered electron projectiles and the host electrons emitted from both sides of the graphene sheet.
By representing electrons both classically, via pseudopotentials, and quantum mechanically, via a Gaussian wave packet (WP), we can isolate the quantum mechanical effect and clearly identify the quantitative difference of those observables with different incident energies.

\section{Computational Methods}

\begin{figure}
    \centering
    \includegraphics[width=0.99\linewidth]{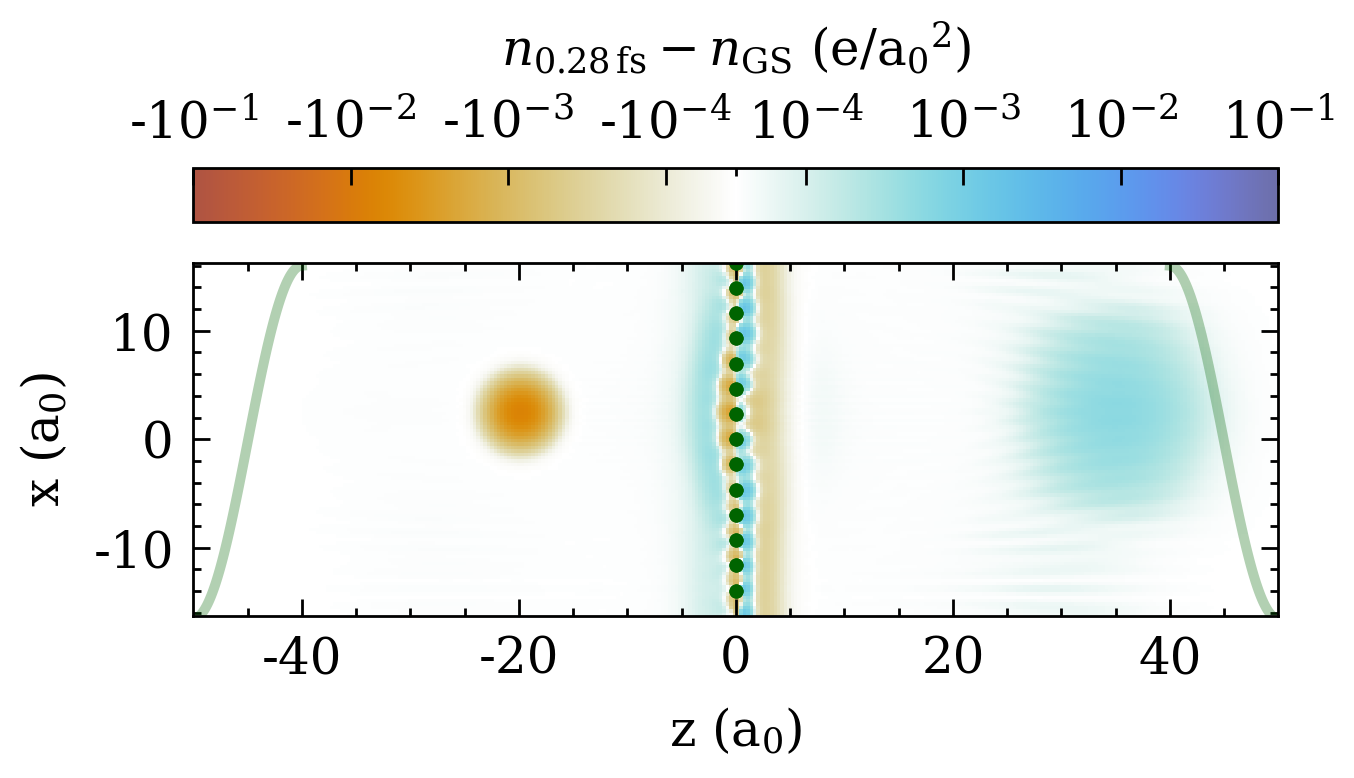}
    \caption{\label{fig:rhoysum}
A schematic of the simulation cell and the planar integrated charge density difference after the electron projectile impacts the graphene (dark green circles represent C atoms) at 0.28 fs after the simulation starts.
The initial wave packet is located at $\left<z\right> = -20\,a_0$ and approaches the graphene perpendicularly.
The solid green line denotes the region of the complex absorbing potential, Eq.~\ref{eq:cap}. 
Two traces of the wave packet are visible, before and after encountering the graphene layer, since this plot shows the electron density \emph{difference}.
}
\end{figure}

Real-time time-dependent density functional theory is used as implemented in the INQ~\cite{andrade_inq_2021} code, to propagate the electrons in the system in time.
Converged ground-state single-particle Kohn-Sham (KS) states from DFT are used as the initial condition of the time-dependent KS equations 
\begin{equation}
    i\frac{\partial}{\partial t} \psi_j(\mathbf{r},t)=\left(-\frac{\nabla^2}{2}+V_{\mathrm{KS}}(\mathbf{r},t)\right)\psi_j(\mathbf{r},t).
\end{equation}
Here, $\psi_j$ are the single-particle KS orbitals evolving in a time-dependent effective potential $V_{\mathrm{KS}}$, which is a functional of the electron density.
KS states are represented as a plane-wave basis with a cutoff energy of 50 Hartree.
Exchange and correlation effects are treated with the adiabatic local density approximation~\cite{zangwill_resonant_1980,zangwill_resonant_1981}.
The electron-ion interaction is described by norm-conserving ONCV pseudopotentials~\cite{hamann_optimized_2013}.
The large simulation cell with 112 carbon atoms and 100 $\mathrm{a_0}$ of vacuum, see Fig.\ \ref{fig:rhoysum}, allows using only the $\Gamma$ point for Brillouin zone sampling.
The maximum force on any of the carbon atoms in the ground state is 3.74 $\mathrm{meV/\AA}$. %
The carbon ions and classical electron, as well as the carbon ions and the wave packet, are propagated following the Ehrenfest dynamics.

In this work, we express the initial state of the incident quantum mechanical electron as a Gaussian wave packet (WP), 
\begin{equation}
\label{eq:wp}
    \psi^{\mathrm{WP}}(\mathbf{r}) = \left(\frac{1}{\pi d^2}\right)^{\frac{3}{4}} e^{-\frac{(\mathbf{r} - \mathbf{b})^2}{2d^2} + i\mathbf{k}\cdot\mathbf{r}},
\end{equation}
where $\mathbf{b}$ and $\mathbf{k}$ are the WP's center position and wave vector, respectively.
The parameter $d$ determines the standard deviation of the position of the WP by $\sigma_r = d/\sqrt{2}$.  
In the main text, we set $d=1.1\,\mathrm{\AA}$, while in Sec.~\ref{sec:wp std} we analyze the electron dynamics arising from different $d$ values ranging from $0.6$ to $1.6\,\mathrm\AA$, and its implications on thicker materials such as multilayer graphene or even bulk graphite.
After the initial state, the propagation of Eq.\ \eqref{eq:wp} is governed by the KS equation.

The total electron density $n(\mathbf{r})$ that determines the KS Hamiltonian is the sum of the electron density of the host material $n^\mathrm{host}$ and the density of the wave packet $\left|\psi^{\mathrm{WP}}(\mathbf{r})\right|^2$,
\begin{equation}
n(\mathbf{r}) = n^{\mathrm{host}}(\mathbf{r}) + \left|\psi^{\mathrm{WP}}(\mathbf{r})\right|^2.
\end{equation} 

The initial position of this electron is centered at 20 $a_0$ from the graphene sheet, as shown in Fig.~\ref{fig:rhoysum}. 
This ensures that the WP at the beginning of the simulation is orthonormal to all Kohn-Sham states of the graphene due to the absence of spatial overlap with the graphene's Kohn-Sham orbitals.
The electron traverses the simulation cell perpendicularly towards the graphene, where its center position follows either a channeling (A) or centroid (O) trajectory, see Fig.~\ref{fig:traj}. 
We use a complex absorbing potential (CAP) to prevent the nonphysical reentering of the electrons into periodic images of the simulation cell.
The details of setting the CAP are provided in Sec.~\ref{sec:cap} of the SI.

For comparison, a classical electron projectile is represented by a pseudopotential; see Sec.~\ref {sec:el psp} for detailed validation of the electron pseudopotential.
To further connect the classical and quantum mechanical approaches, we also prepare an ensemble of classical electrons, whose initial positions and momenta are sampled from the positional and momentum distributions of the Gaussian WP, see Sec.~\ref {sec:clel dist} for the details of generating the classical electron ensemble.

\section{Results}

To systematically study both the quantum mechanical and classical regimes, we select the kinetic energies of the incident electrons based on their corresponding de Broglie wavelengths.
For this, we simulate incident electrons with a kinetic energy range from 100 eV to 800 eV, corresponding to de Broglie wavelengths of about 1.23 $\mathrm{\AA}$ to 0.43 $\mathrm{\AA}$.
This is comparable to the length of the C--C bond in graphene, 1.42 $\mathrm{\AA}$, suggesting that quantum mechanical effects such as diffraction, tunneling, and reflection are significant at these energies or below.
A wave packet with a kinetic energy of 7.4 keV has a de Broglie wavelength of 1/10 of the C--C bond in graphene, indicating that a classical description of the electron projectile may be sufficient.
The significance of quantum mechanical effects will be highlighted later when we discuss the corresponding observables.

\begin{figure}
    \centering
    \includegraphics[width=0.95\linewidth]{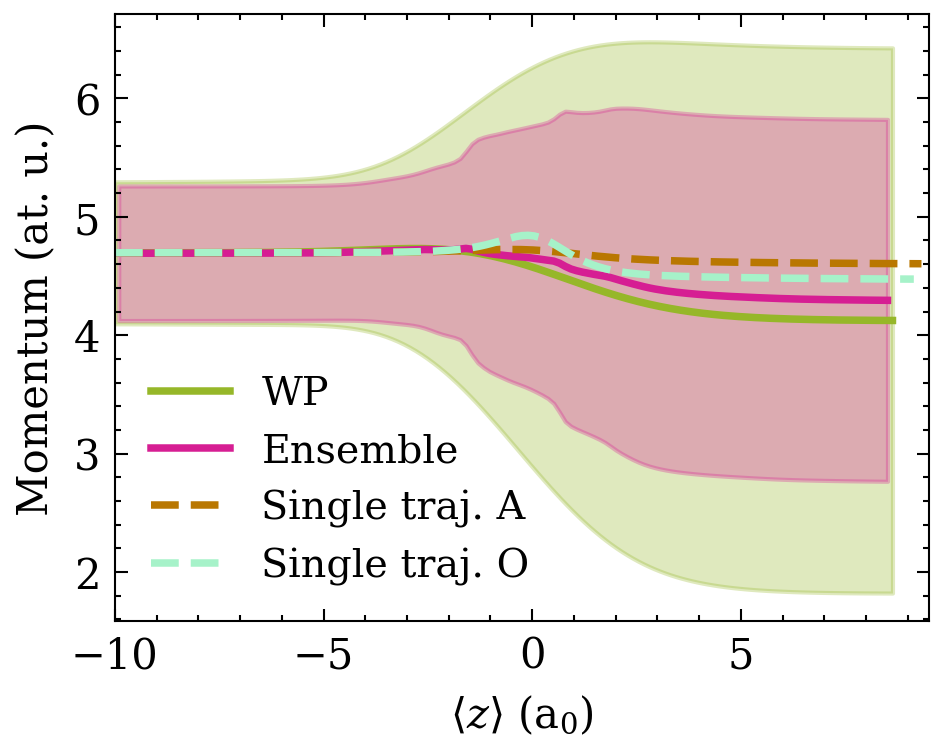}
    \caption{\label{fig:clel 300 dist v}
    Comparison of the momentum distribution of electrons as a function of the mean position $\left<z\right>$ for the wave packet (WP) versus classical approaches with the mean velocity corresponding to 300 eV. 
    The WP and ensemble of classical trajectories are centered at the channeling trajectory, Traj.\ A.
    We also include one additional centroid trajectory, Traj.\ O.
    The solid lines and shaded regions denote the mean and the standard deviation of the momentum, respectively.
    The impact point of the Traj.\ A and Traj.\ O is marked in Fig.~\ref{fig:traj}.
    }
\end{figure}

We first investigate the momentum distribution 
of electron projectiles as they irradiate onto graphene:
Figure~\ref{fig:clel 300 dist v} compares the momentum distribution of the WP, a single classical electron trajectory, and an ensemble of classical electrons as a function of position.
The single channeling trajectory exhibits a smaller decrease in momentum compared to both the WP and classical ensembles. This arises because channeling through the hexagonal center avoids high electron‐density regions near C–C bonds, reducing electronic stopping.
The single \emph{centroid} trajectory exhibits a greater decrease in momentum after impact compared to the single \emph{channeling} trajectory, as its impact point is closer to the directional bond, as shown in Fig.~\ref{fig:traj}. 
Conversely, the classical electron ensemble aligns more closely with the quantum mechanical WP in both the mean and standard deviation of the momentum. 
This similarity occurs because both the wave packet and the classical ensemble interact with graphene over a spatially extended region, unlike the point-like interaction of a single classical trajectory.

\begin{figure}
    \centering
    \includegraphics[width=0.99\linewidth]{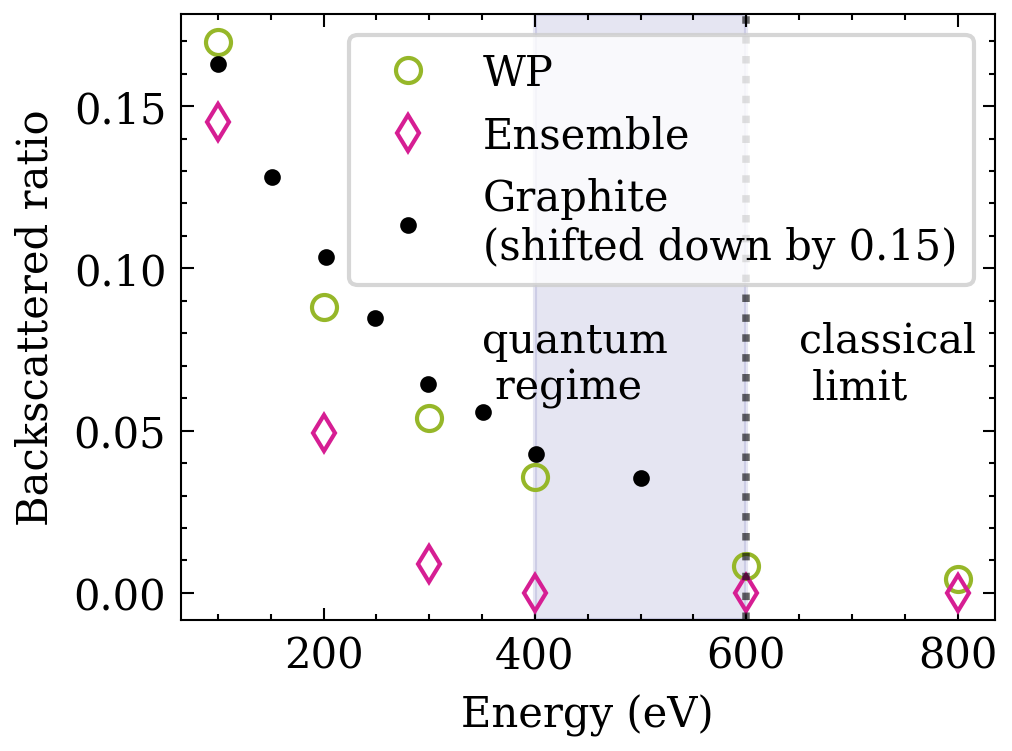}
    \caption{\label{fig:be com}
    The ratio of the backscattered electrons to the total incident electrons simulated from the wave packet (WP), and an ensemble of classical trajectories, compared with the experimental backscattered yield for a bulk graphite~\cite{farhang_electron_1993}, shifted down by a constant 0.15. 
    The vertical black dashed line separates where the quantum mechanical effect vanishes.
    The region where the classical effect vanishes and the purely quantum regime appears is shaded in purple.  
    }
\end{figure}

The evolution of the momentum standard deviation is qualitatively similar between the WP and the classical electron ensemble, see Fig.\ \ref{fig:clel 300 dist v}.
Before impact, the standard deviation remains constant, as expected for particles propagating freely in a vacuum.
Upon scattering with graphene, momentum randomization increases the standard deviation for both the classical ensemble and the quantum-mechanical WP.
However, the underlying physics is distinct: 
Each classical electron in the classical ensemble follows a deterministic trajectory based on the Ehrenfest forces from the interaction potential, and the standard deviation reflects the statistical spread of these independent paths.
In contrast, for the quantum mechanical WP, the increase of the momentum standard deviation arises from genuine wave phenomena, such as diffraction, tunneling, and reflection, which result in a less uniform momentum distribution:
One direct consequence of these quantum effects is the non-zero backscattered electron yield at incident energies of 400 eV, which we will examine further in Fig.~\ref{fig:be com}.

\begin{figure}
    \centering
    \includegraphics[width=0.99\linewidth]{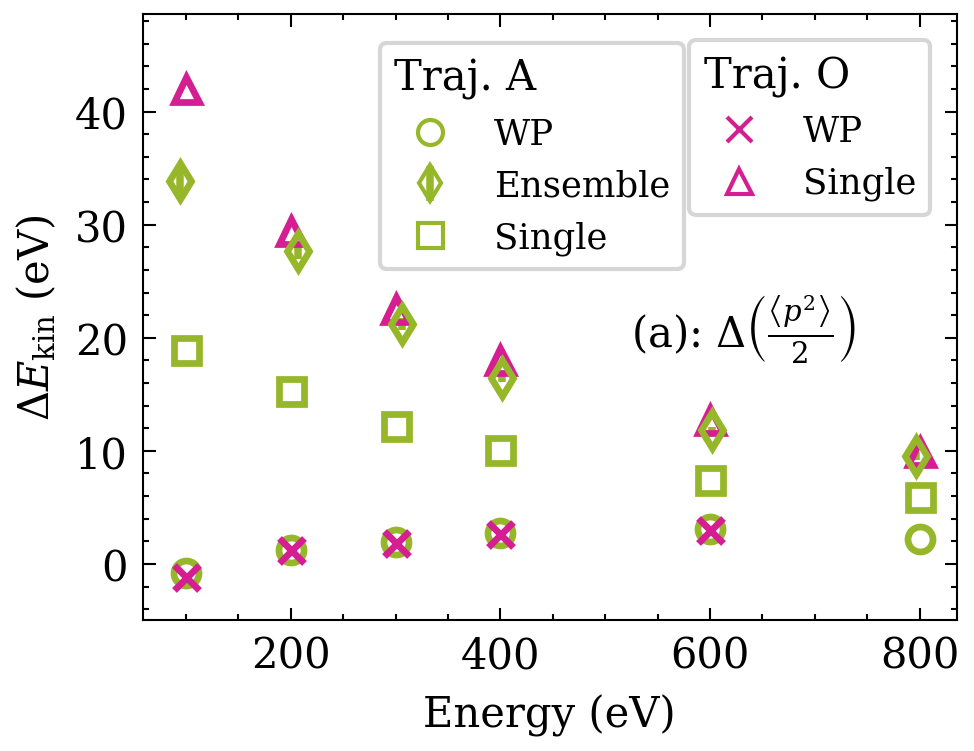}
    \includegraphics[width=0.99\linewidth]{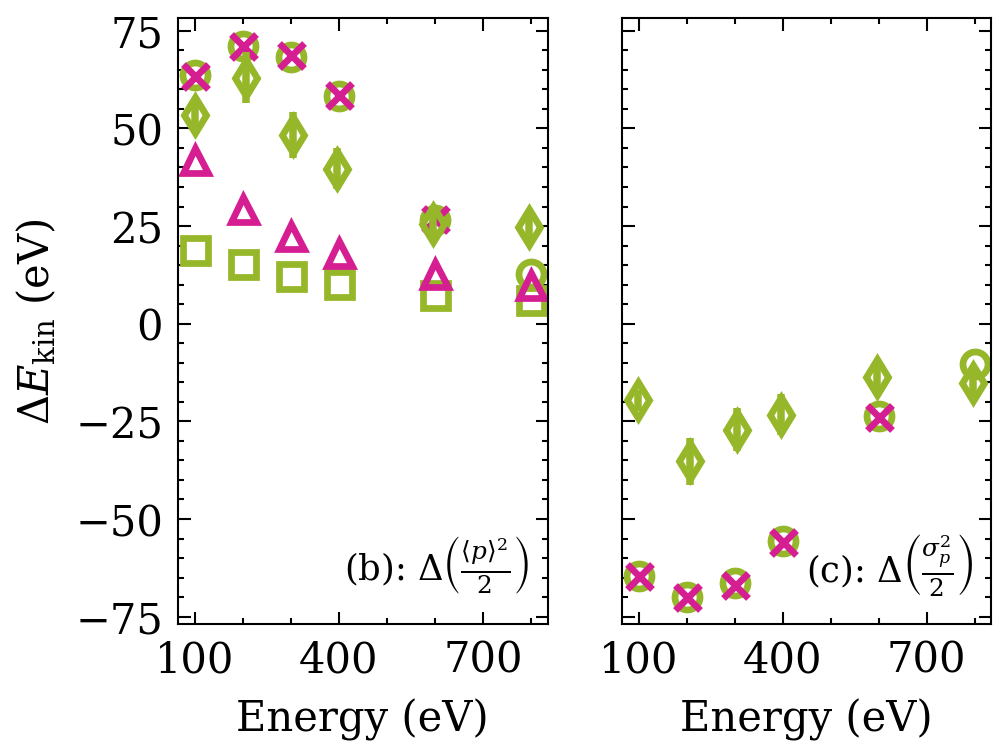}
    \caption{\label{fig:dKE}
    The kinetic energy loss of electrons described as a quantum mechanical wave packet (WP), an ensemble of classical electrons, or a single classical electron, centered on two distinct trajectories: 
    channeling (Traj. A) and centroid (Traj. O).
    The total kinetic energy loss, $\Delta E_{\text{kin}}$, calculated from Eq.~\eqref{eq:Ekin} and plotted in panel (a), is decomposed into (b) contributions from the average momentum, \(\Delta \left( \left< p \right>^2 / 2 \right)\), and (c) the momentum spreading, \(\Delta \left( \sigma_p^2 / 2 \right)\).
    The error bar in the classical ensemble indicates the standard error of the mean of the kinetic energy loss.
    }
\end{figure}

The evolution of the momentum distribution directly quantifies the kinetic energy loss of the incident electron, which can be calculated by
\begin{equation}
\begin{split}
    E_\mathrm{kin} &= \frac{\left<p^2\right>}{2} = \frac{\left<p\right>^2}{2} + \frac{(\sigma_p)^2}{2}, \\ 
    \Delta E_\mathrm{kin} &= E_\mathrm{kin}\left(\left<z\right> = -20\, a_0\right) - E_\mathrm{kin}\left(\left<z\right> = 20 \, a_0\right).
\end{split}
 \label{eq:Ekin}
\end{equation}
The average momentum $\left<p\right>$ and the standard deviation of the momentum $\sigma_p$ can be extracted from Fig.~\ref{fig:clel 300 dist v}.
The kinetic energy loss is then the difference between the kinetic energy before impact, $\left<z\right> = -20\, a_0$, and after impact, $\left<z\right> = 20 \, a_0$.
We further decompose the kinetic energy loss into the change in mean momentum, $\Delta \left(\left< p \right>^2/2\right)$, Fig.~\ref{fig:dKE}\,b, and the change in momentum spreading, \(\Delta \left(\sigma^2_p/2\right)\), in Fig.~\ref{fig:dKE}\,c. 
These two components exhibit competing effects:
scattering against graphene randomizes the momentum of the incident electrons, resulting in a decrease in the mean momentum, see the positive value in Fig.~\ref{fig:dKE}\,b, and an increase in the momentum standard deviation, see the negative value in Fig.~\ref{fig:dKE}\,c. 
This is consistent with the real-time momentum distribution of the WP and the classical ensemble in Fig.~\ref{fig:clel 300 dist v}.
When comparing the classical electron ensemble to the quantum WP, the classical model successfully captures the qualitative behavior of both individual components, shown in Fig.~\ref{fig:dKE},b and c.
However, because the total kinetic energy loss is the sum of these two competing, oppositely signed terms, the final result is highly sensitive to their exact magnitudes.
As shown in Fig.~\ref{fig:dKE}\, a, the sum of these two terms yields a distinct trend in the kinetic energy loss of the classical electron ensemble compared to that of the WP.
This indicates that while the scattering effects, such as slowdown and broadening, are qualitatively captured by the classical ensemble, the precise quantitative kinetic energy loss still requires the quantum mechanical description.

For classical electrons described by a single trajectory, the kinetic energy loss exhibits a sensitive trajectory dependence.
This results from the different charge densities along various trajectories due to the directional bonding in graphene, as depicted in Fig.~\ref{fig:traj}. 
The sensitive trajectory-dependent kinetic energy loss is consistent with the light-ion irradiation of graphene~\cite{kononov_anomalous_2021, kononov_first-principles_2022}. 
In contrast, the kinetic energy loss of the WP, whether centered on the channeling or centroid trajectory, remains similar, as shown in Fig.~\ref{fig:dKE}\,a. 
This similarity arises because the real-space spreading of the WP inherently samples the graphene lattice, reducing trajectory-specific effects. 
Hence, unless specified, we only focus on the WP centered at the channeling trajectory in the remainder of the text.

The centroid trajectory approximates the ensemble-averaged kinetic energy loss, as shown in Fig.~\ref{fig:dKE}\,a.
This confirms its effectiveness in representing ensemble properties not only for light ion irradiation in various materials~\cite{ojanpera_electronic_2014, yost_electronic_2016}, but also for classical electron irradiation. 
Interestingly, Fig.~\ref{fig:dKE}\,(b) shows that the change of the mean momentum in the classical ensemble is not the same as the single centroid trajectory (Traj.\ O). 
This indicates that the centroid trajectory inherently encodes information about both the mean and standard deviation of the momentum of the classical ensemble.

As the electron emission may last around two fs after the electron impacts the material, estimated from proton-irradiated graphene~\cite{yao_nonequilibrium_2024}, simulating the classical ensemble with more than 100 trajectories for such a long time is computationally impractical using rt-TDDFT. We will hence rely on the centroid trajectory for the ensemble-averaged electron emission dynamics. 
    
\begin{figure}
    \centering
    \includegraphics[width=0.99\linewidth]{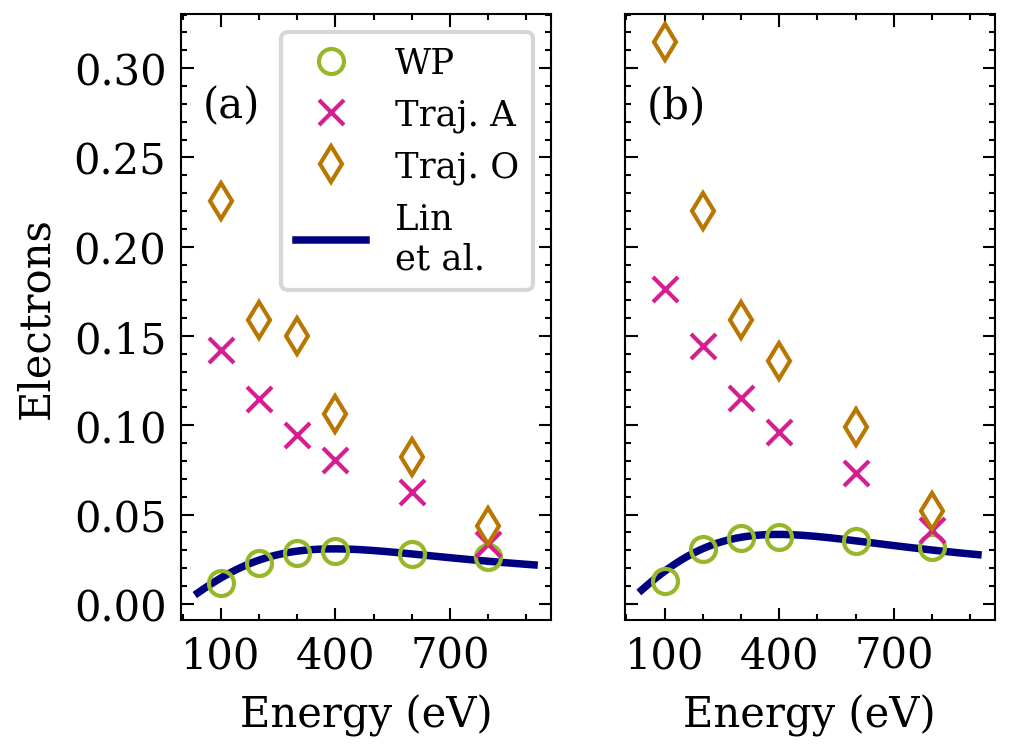}
    \caption{\label{fig:sec com}Secondary electron emission at (a) the entrance side and (b) the exit side of graphene irradiated by: a wave packet (WP) centered at the channeling trajectory (Traj. A), a single classical electron along the channeling trajectory (Traj. A), and a single classical electron along the centroid trajectory (Traj. O). 
    The solid line represents the universal law for secondary electron yield derived by Lin \emph{et al.}\ in Ref.~\cite{lin_new_2005}, see Sec.~\ref{sec: universal law of sey} in the SI for details of how we calculate this curve.}
\end{figure}

After the energy of the projectile is deposited into the graphene, the excited electrons can potentially overcome the work function and emit into the vacuum, known as secondary electrons.
These secondary electrons can then be used for imaging \cite{seiler_secondary_1983} and potentially probe the lattice temperature in the material \cite{yao_nonequilibrium_2024}. 
For the WP, the electron emission exhibits a similar trend on both sides of the graphene. However, the maximum yield on the exit side, for 400 eV, is approximately 30\,\% higher than that on the entrance side, see green circles in Fig.~\ref{fig:sec com}.
This exhibits a similar trend and maximum of the simulated secondary electron emission of WP-irradiated graphene flakes \cite {ueda_quantum_2016}.
The direct experimental comparison remains challenging because secondary electron emission from a single-layer graphene is difficult to isolate from substrate contributions~\cite{bundaleska_prospects_2020,wang_secondary_2016,feng_characteristics_2022}. 
However, empirical models based on binary scattering approximations capture the general shape of secondary electron yield versus incident energy~\cite{lin_new_2005}.
The solid line in Fig.~\ref{fig:sec com} represents this universal secondary electron yield law~\cite{lin_new_2005}, which has shown good agreement across various solids~\cite{walker_secondary_2008, mehnaz_exploring_2023} (see SI Sec.~\ref{sec: universal law of sey} for details on employing this model in our simulations).
While this empirical model cannot provide quantitative predictions like our rt-TDDFT simulations, it successfully reproduces the general shape of the secondary electron yield curve on both sides of the WP-irradiated graphene.

A fraction of the incident electrons is reflected back from the target material, known as backscattered electrons.
Figure~\ref{fig:be com} illustrates the ratio of backscattered electrons at the entrance side vacuum, comparing results from a WP, an ensemble of classical trajectories, and an empirical model for bulk graphite (shifted down by a constant of 0.1). 
As expected, the absolute number of backscattered electrons is more significant at lower incident kinetic energies in both the WP and classical ensembles. 
However, the distinction between the classical ensemble and the WP becomes significant as the incident energy increases. Around 400 eV, the classical ensemble's backscattered ratio drops to zero, whereas the WP maintains a finite value, e.g., around 0.04 at 400 eV. 
This nonzero backscattering in the WP highlights a quantum-mechanical effect that cannot be captured by an ensemble of classical electrons.
Consequently, this intermediate kinetic-energy regime provides a possible experimental access for isolating quantum-mechanical backscattering processes with minimal classical contributions.
Such kinetic energy regime might be relevant to phase-sensitive measurement such as coherent backscattering~\cite{berry_coherent_1994}, or even spin-dependent scattering technique, such as spin-polarized low energy electron diffraction~\cite{r_feder_spin-polarised_1981}. 
To observe a comparable quantum-mechanical backscattering effect of an ion, the projectile ion must possess a similar de Broglie wavelength. A proton would requires an ultra-low kinetic energy of approximately 0.22 eV, at which the charge transfer or surface adsorption alter the interaction potential and the resulting scattering process, which still make the electron a more favorable probe of these quantum mechanical effects. 
Besides, this higher backscattered ratio from the WP indicates a less uniform motion, which is consistent with the higher momentum spreading shown in Fig.~\ref{fig:clel 300 dist v} and Fig.~\ref{fig:dKE}~c.
As the kinetic energy of the incident electrons increases further, e.g.\ at an incident energy of 800 eV, the backscattered ratio from WP approaches 0, indicating that the quantum mechanical effect diminishes.
The corresponding de Broglie wavelength is 0.43 $\mathrm{\AA}$, which is about 30\,\% of the C-C bond in graphene.  

Although no direct experimental data for the backscattered ratio of graphene is available, our results still exhibit a similar decreasing trend compared to the experimental data for the backscattered electron yield of bulk graphite~\cite{farhang_electron_1993}, see Fig.~\ref{fig:be com}.
Our predicted yield for graphene is smaller and declines more rapidly than the experimental bulk graphite data. 
This difference arises because graphene's single-atom thickness provides fewer scattering centers for electron backscattering compared to bulk graphite. 
The similar trend, despite the quantitative difference, underscores the role of material dimensionality in electron scattering, while validating the qualitative reliability of our simulated backscattering behavior.
Further simulational and experimental comparison is still being sought.

\begin{figure}
    \centering
    \includegraphics[width=0.99\linewidth]{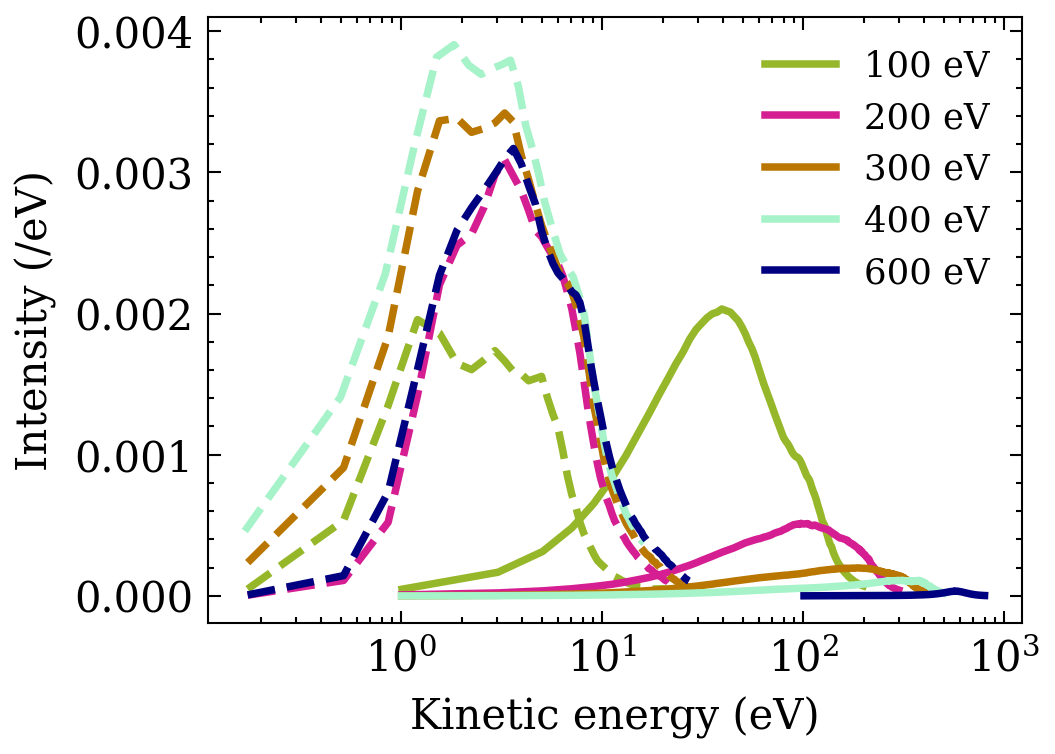}
    \caption{
\label{fig:ke com}
The kinetic energy distribution of electrons in the entrance side vacuum is decomposed into the backscattered electrons from the wave packet (solid lines) and the secondary electrons (dashed lines).
The curves are normalized to the secondary-electron yield or the backscattered-electron yield shown in Fig.~\ref{fig:be com}.
}
\end{figure}

The kinetic energy distributions of secondary electrons from graphene and backscattered electrons from the WP are of interest in electron–matter interaction studies.
Secondary electrons are generated by inelastic collisions in which the incident electron transfers a small fraction of its energy to the bound electron of the host material. Hence, the kinetic energy spectrum is narrow and peaked at low energies, typically below ~50 eV, with a dominant peak around a few eV~\cite{seiler_secondary_1983,bouchard_surface_1980}.
The distinction of the secondary electrons emitted from the host material and backscattered electrons from the incident wave packet is unambiguous in our simulation. 
The kinetic energy spectra are computed using the same methodology described in Ref.~\cite{yao_nonequilibrium_2024}.
The dashed lines in Fig.~\ref{fig:ke com} reveal that the kinetic energy distribution of the secondary electrons from the graphene is narrow and peaked at a few eV, regardless of the energy of the incident electrons. 
Conversely, the peak of the backscattered electrons shifts toward higher energies as the incident electron energy increases, see solid curves in Fig.~\ref{fig:ke com}, retaining a significant fraction of the initial energy.

This spectral separation becomes less distinct at lower incident energies. 
For example, with a 100 eV incident electron, backscattered electrons can have kinetic energies that partially overlap with the range of secondary electrons, see the green solid and dashed lines in Fig.~\ref{fig:ke com}.
Despite this spectral overlap, our rt-TDDFT results reveal that the yields of these two processes differ by an order of magnitude in this regime. 
For a 100 eV incident electron, the backscattered electron yield of 0.169, see Fig.~\ref{fig:be com}, is more than an order of magnitude higher than the secondary electron yield of 0.01, see Fig.~\ref{fig:sec com} a. 
This demonstrates that even when their kinetic energy distributions partially overlap, one emission process dominates the total yield at a given incident energy.
Specifically, backscattering dominates at low incident energies, while secondary emission becomes dominant as the incident energy increases. 

\section{Conclusion}
In conclusion, we employed real-time time-dependent density functional theory to simulate the electron dynamics under external electron irradiation.
This work provides unprecedented insight into the femtosecond dynamics of momentum evolution and the kinetic energy loss of the incident electrons, comparing both classical and quantum mechanical descriptions. 
Furthermore, we quantitatively predict the experimentally elusive secondary and backscattered electron characteristics of freestanding single-layer graphene, which require experimental verification isolated from substrate interference.  
Within the 300–600 eV incident energy range, we identify a clear transition regime where backscattered electron yields from a classical point-charge projectile diverge from those of a quantum wave-packet description. 
The quantitative differences observed in kinetic energy loss, secondary electron yield, and backscattered electron yield between classical point-charge and quantum wave-packet descriptions underscore the critical importance of quantum mechanical effects in electron irradiation.
Our results offer clear guidance for future simulations: classical descriptions may suffice in high-energy regimes where wave-like behavior is negligible. At the same time, quantum treatments are essential in the low-to-intermediate energy range where quantum-mechanical effects become significant.

Beyond theoretical implications, our work establishes a foundation for the rational design of two-dimensional materials under electron irradiation, with direct relevance to nanofabrication processes and high-resolution electron-beam technologies.
Importantly, quantifying secondary and backscattered electron yields is a critical step toward predicting radiation damage, such as sputter coefficients, which govern material modification under electron exposure.
Although simulating kinetic energy loss and emission dynamics for classical ensembles remains computationally demanding, machine-learning strategies such as that proposed in Ref.~\cite{ward_accelerating_2024} may offer a practical route to accelerated predictions.
Moreover, the identified energy regime around 400 eV may provide an experimentally accessible “quantum-only” scattering window for studying, for example, the coherent quantum scattering and spin-polarized low energy electron diffraction in solids.

\begin{acknowledgments}
We thank Nicola Perry, Elif Ertekin, Alfredo Correa, Alina Kononov, Emilio Artacho, Xavier Andrade, Gerhard Hobler, Min Choi, Bryan Wong, P.K. Verma, and Xiuyao Lang for the helpful discussion.
This material is based upon work supported by the National Science Foundation under Grant No.\ OAC-2209857. 
This work made use of the Illinois Campus Cluster, a computing resource that is operated by the Illinois Campus Cluster Program (ICCP) in conjunction with the National Center for Supercomputing Applications (NCSA) and which is supported by funds from the University of Illinois at Urbana-Champaign.
This work used the Delta system at the National Center for Supercomputing Applications through allocation MAT240105 and  MAT240058 from the Advanced Cyberinfrastructure Coordination Ecosystem: Services \& Support (ACCESS) program, which is supported by National Science Foundation grants \#2138259, \#2138286, \#2138307, \#2137603, and \#2138296.
\end{acknowledgments}

\bibliography{lib}

\beginsupplement

\begin{center}
\textbf{Supplemental Material for\\\vspace{0.5 cm}
Electron-irradiated graphene
\\\vspace{0.3 cm}}
Yifan Yao$^{1}$ and Andr\'e Schleife$^{1,2,3}$

\small
$^1$\textit{Department of Materials Science and Engineering, University of Illinois, Urbana-Champaign, Urbana, IL 61801, USA}
$^2$\textit{Materials Research Laboratory, University of Illinois at Urbana-Champaign, Urbana, IL 61801, USA}
$^3$\textit{National Center for Supercomputing Applications, University of Illinois at Urbana-Champaign, Urbana, IL 61801, USA}
\end{center}

\section{Pseudopotential for electrons}
\label{sec:el psp}
We adopt the pseudopotential approach to simulate the classical incident electron.
The pseudopotential of an electron is created by inverting the local part of the pseudopotential of a hydrogen atom, inverting the valence charge to $-1$, and setting the mass to 1 at.\,u..   

\begin{figure}
    \centering
    \includegraphics[width=0.618\linewidth]{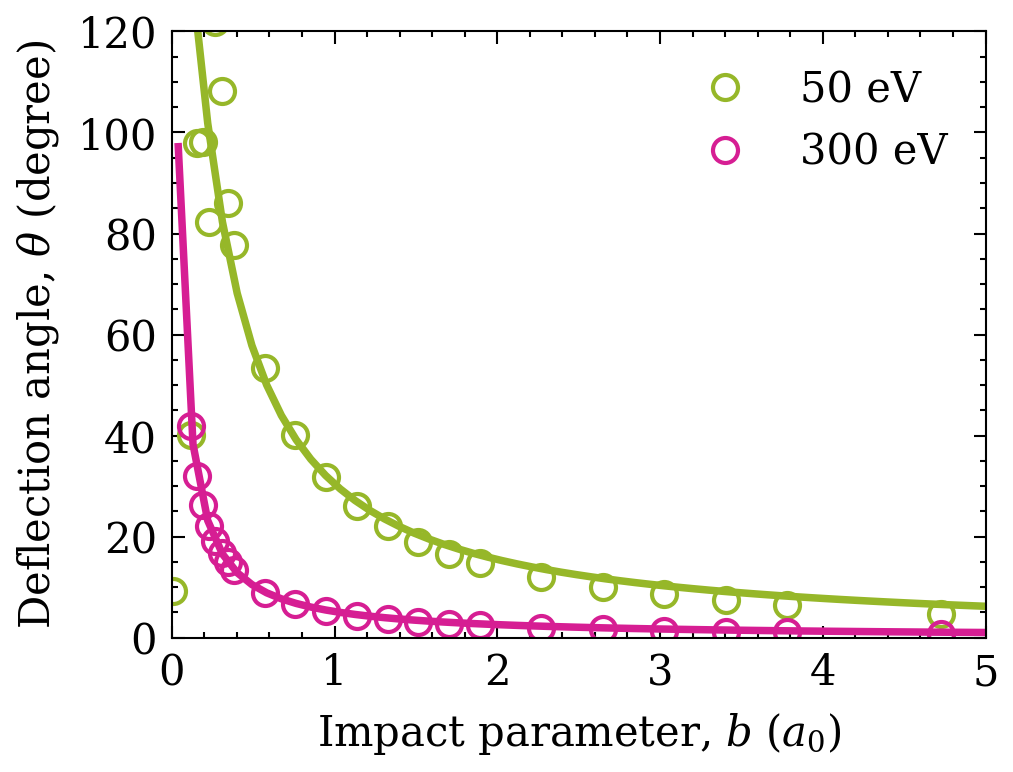}
    \caption{The deflection angle of classical electrons impacting an isolated $\mathrm{H^{0.9+}}$, to mimic the hydrogen nucleus. The solid lines denote the analytical solution of the deflection angle \(\theta\) from the Rutherford model as a function of the impact parameter \(b\), Eq.~\ref{eq:deflection angle}, and the empty circles are from the rt-TDDFT. A good agreement has been observed between the simulated deflection angle and the analytical one.}
    \label{fig:rutherford}
\end{figure}

To validate our pseudopotential for an incident classical electron, we irradiate the electron onto an isolated $\mathrm{H^{0.9+}}$ to mimic the hydrogen nucleus. We compare the deflection angle with the Rutherford model, which is derived from the bare Coulomb potential. Fig.~\ref{fig:rutherford} shows the deflection angle of the incident electron along with the Rutherford model written in the atomic unit for the incident electron, 
\begin{equation}
    \tan\left(\frac{\theta}{2}\right) = \frac{Z_1Z_2}{2b\cdot E_\mathrm{kin}}~,
    \label{eq:deflection angle}
\end{equation}
where $\theta$ denotes the deflection angle, $b$ denotes the impact parameter, and the $E_\mathrm{kin}$ denotes the kinetic energy of the incident electrons. The charge state of the incident electron and the host ions are denoted by $Z_1$ and $Z_2$.
The deflection angle calculated by the rt-TDDFT and Ehrenfest dynamics agrees excellently with the Rutherford model, which validate our pseudopotential for the classical electrons. 

\section{Trajectories}
\begin{figure}
    \centering
    \includegraphics[width=0.618\linewidth]{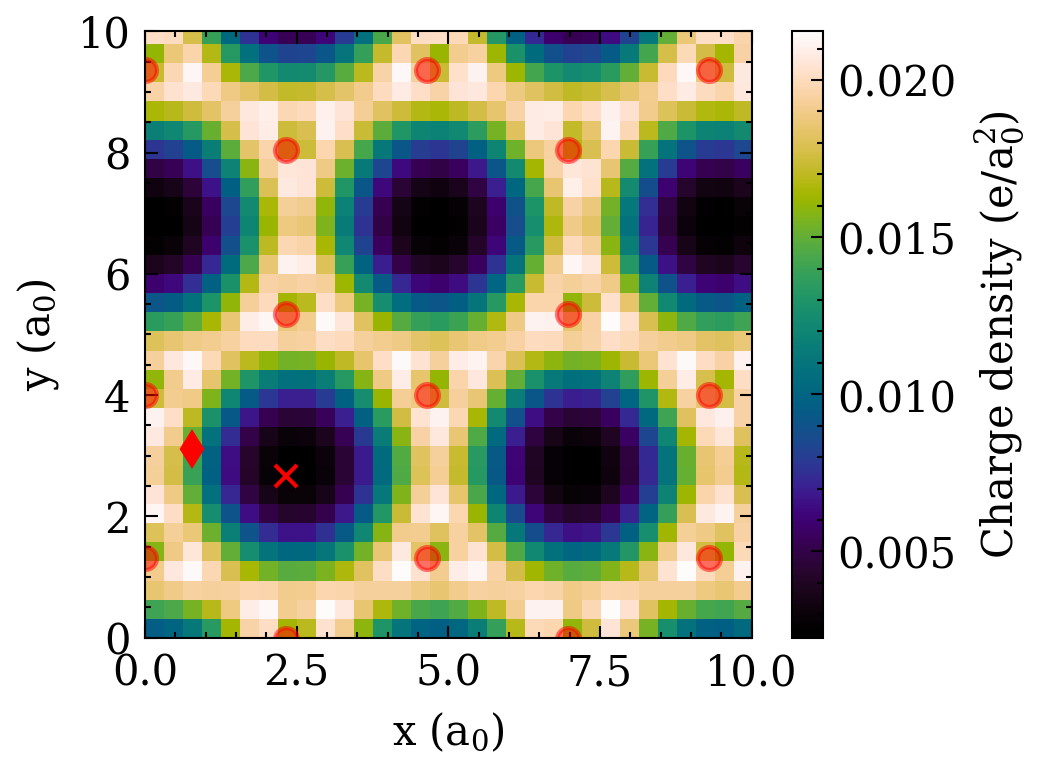}
    \caption{The ground state charge density of graphene overlaid with the carbon atom (red circle), the channeling trajectory, traj. A (red $\mathrm{X}$), and the centroid trajectory, traj. O (red diamond)}.
    \label{fig:traj}
\end{figure}

Two trajectories of the electrons impacting the graphene are examined in this work: their center position follows two different trajectories, channeling trajectory (Traj. A) and centroid trajectory (Traj. O). 
Fig.~\ref{fig:traj} shows the center of the trajectories, along with the atoms and the integrated charge density of the targeted graphene.

\section{Absorbing potential for wave packet with different incident velocity}
\label{sec:cap}
As the wave packet traverses the host materials, electrons are emitted from both sides of the host materials and propagate outwards.
We employ a complex absorbing potential to avoid the non-physical re-approach of the wave packet and the emitted electrons to the graphene due to the periodic boundary condition. It has a form of~\cite{kononov_electron_2022, de_giovannini_modeling_2015} 
\begin{equation}
    \mathrm{V}_{\mathrm{CAP}}(z)=-i\cdot \mathrm{W}\cdot \sin^2\left(\frac{(z-\mathrm{z_s})\cdot \pi}{2\cdot dz}\right)~,
    \label{eq:cap}
\end{equation}
for $\mathrm{z_s}<z<\mathrm{z_s}+2dz$, where $\mathrm{z_s}$ is the front boundary of CAP, and $dz$ is the half width of the CAP. 
In our simulation, we set $\mathrm{z_s}=0.5$, fix the width, $dz=0.1$ of the absorbing potential, and tune the height, $\mathrm{W}$.

To find out the most suitable $\mathrm{W}$ for each incident wave packet, we compare the remaining electrons of the wave packet in a vacuum, without graphene, after impacting the absorbing potential. 
We select the $\mathrm W$ that gives the smallest remaining electrons in the vacuum. 
In the actual simulation of irradiated graphene, the wave packet will be distorted after interacting with the graphene, which may alter the optimal $\mathrm{W}$. However, the test here still provides a reasonable estimation of the value of $\mathrm{W}$. 

\begin{figure}
    \centering
    \includegraphics[width=0.618\linewidth]{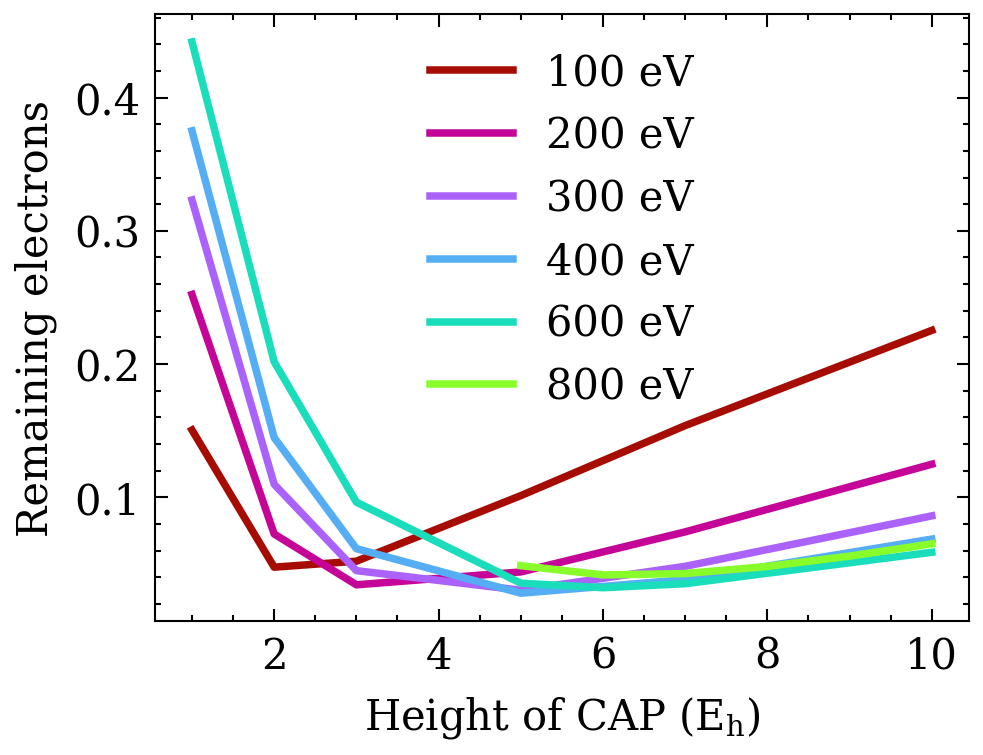}
    \caption{Remaining electrons in the vacuum after the wave packet reaches the complex absorbing potential.}
    \label{fig:CAPWtune}
\end{figure}

\section{Ensemble of classical electrons}
\label{sec:clel dist}
We use an ensemble of classical trajectories to approximate the quantum mechanical behavior. 
As the Gaussian wave packet remains the Gaussian distribution in both position and momentum, we use \texttt{std::mt19937} in \texttt{C++} to sample Gaussian random numbers in both position and momentum space, generated from the same mean and standard deviation as the corresponding Gaussian wave packet. The input files are included in the \yy{mdf}. 
More than 100 classical trajectories in the ensemble at each kinetic energy. 

\section{Position and momentum distribution of the wave packet}
The position of the wave packet can be calculated as 
\begin{equation}
    \left<\mathbf{r}\right> = \int \mathbf{r} \left|\psi(\mathbf{r})\right|^2 d\mathbf{r}~,
\end{equation}
where $r$ denotes the Cartesian coordinates. 
However, the mean position becomes ill-defined when the wave function crosses the periodic boundary. 
Here, we use the circular mean to calculate the expected positions of the wave packet. 

\begin{align}
    \begin{split}
        z &= e^{2\pi i \tilde{\mathbf{r}}} \\
        \left<z\right> &= \int z \left|\psi(\mathbf{r})\right|^2 d\mathbf{r}\\
        \theta &= \mathrm{Arg} \left(\left<z\right>\right) \\
        \left<\tilde{\mathbf{r}}\right> &= \frac{\theta}{2\pi}
    \end{split}
\end{align}
where $\tilde{\mathbf r}$ is the contravariant coordinate in the crystal, which can then be converted to the  Cartesian coordinate, $\mathbf r$.

Then, we calculate the variance of the position of the wave packet.
\begin{equation}
    \mathbf{\sigma}^2_r = \int \left(\mathbf{r} - \left<\mathbf{r}\right>\right)^2 \left|\psi(\mathbf r)\right|^2 d\mathbf r~,
\end{equation}
where we take the minimum image convention for the $\mathbf{r} - \left<\mathbf{r}\right>$.

The expected momentum distribution is calculated,
\begin{equation}
    \left<\mathbf p\right> = \int \mathbf{p} \left|\phi(\mathbf{p})\right|^2 d\mathbf{p},
\end{equation}
where $\phi(\mathbf{p})$ is the Fourier transform of the real space wave function $\psi(r)$ of the wave packet.

Similarly, we calculate the variance of the momentum, 
\begin{equation}
    \mathbf{\sigma}_p^2 = \int \left(\mathbf{p}-\left<\mathbf{p}\right>\right)^2 \left|\phi(\mathbf{p})\right|^2 d\mathbf{p}~.
\end{equation}

For the Gaussian wave packet, the standard deviation of the momentum is related to the standard deviation of the position by $\sigma_p = \frac{1}{2\sigma_r}$, which is used as a unit test for the implementation. 

\begin{figure}
    \centering
    \includegraphics[width=0.618\linewidth]{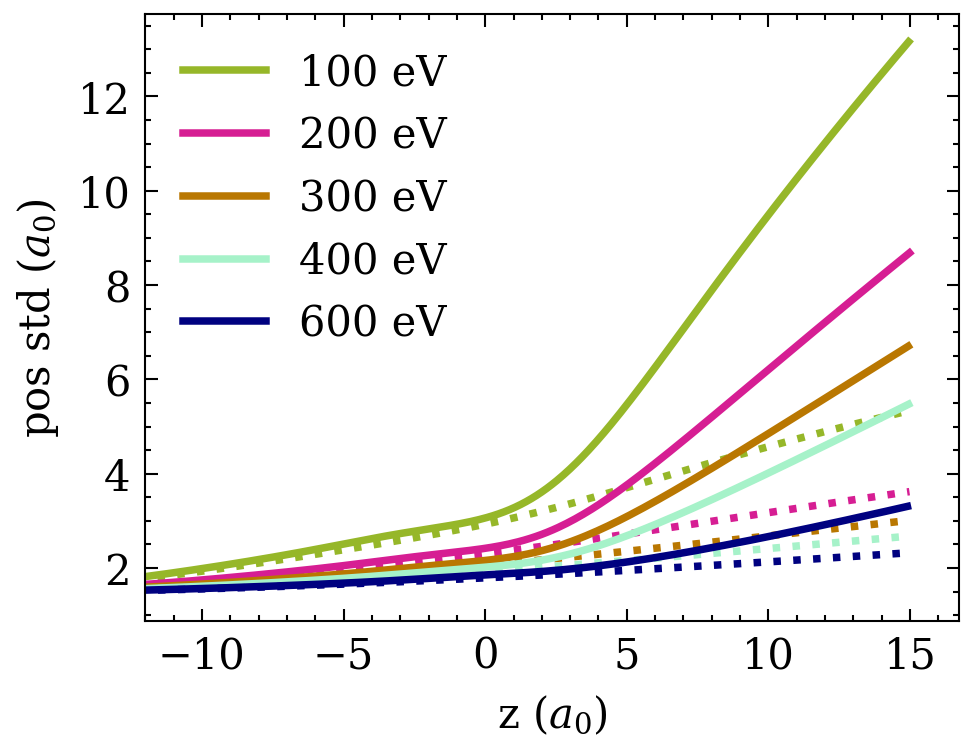}
    \caption{The standard deviation of the position of the wave packet as a function of time. The dotted lines denote the analytical expression of the wave packet position standard deviation as it evolves in the vacuum.}
    \label{fig:wp std com}
\end{figure}

In real-time propagation, the position standard deviation of a wave packet that evolves in free space follows 
\begin{equation}
    \sigma_r(t) = \sigma_r(t=0) \sqrt{1 + \left( \frac{t}{2\sigma_r^2(t=0)} \right)^2},
    \label{eq:vac wp std}
\end{equation}
which can be used to verify our implementation before the wave packet impacting the graphene, and the deviations from this analytical solution after impact, as illustrated in Fig.~\ref{fig:wp std com}.

\section{Universal law of the secondary electron yield}
\label{sec: universal law of sey}
We follow the equation derived in Ref.~\cite{lin_new_2005} to get a universal law of the secondary electron yield, and compare that with the yield from the wave packet irradiation. 
\begin{equation}
    \delta = \delta^\mathrm m \cdot 1.28 \left(\frac{E}{E^\mathrm m}\right)^{-0.67} \cdot \left( 1-e^{-1.614 \cdot \left(\frac{E}{E^\mathrm m}\right)^{1.67}} \right) ~,
\end{equation}
where the $E$ is the energy of the incident electrons, and $E^\mathrm{m}$ is the energy of the incident electrons at which the maximum secondary electron yield occurs. In our case, it is 400 eV. 
$\delta$ is the predicted secondary electron yield, and the $\delta^\mathrm{m}$ is the maximum secondary electron yield from the wave packet irradiation, which in our case is 0.029 in the entrance side and 0.0377 in the exit side. 
While the equation is derived for the backward emitted electron, we apply it to the forward emitted electrons in the exit side vacuum, since our sample is a single-layer graphene.  

\section{Initial state of wave packet} \label{sec:wp std}

\begin{figure}
    \centering
    \includegraphics[width=0.46\linewidth]{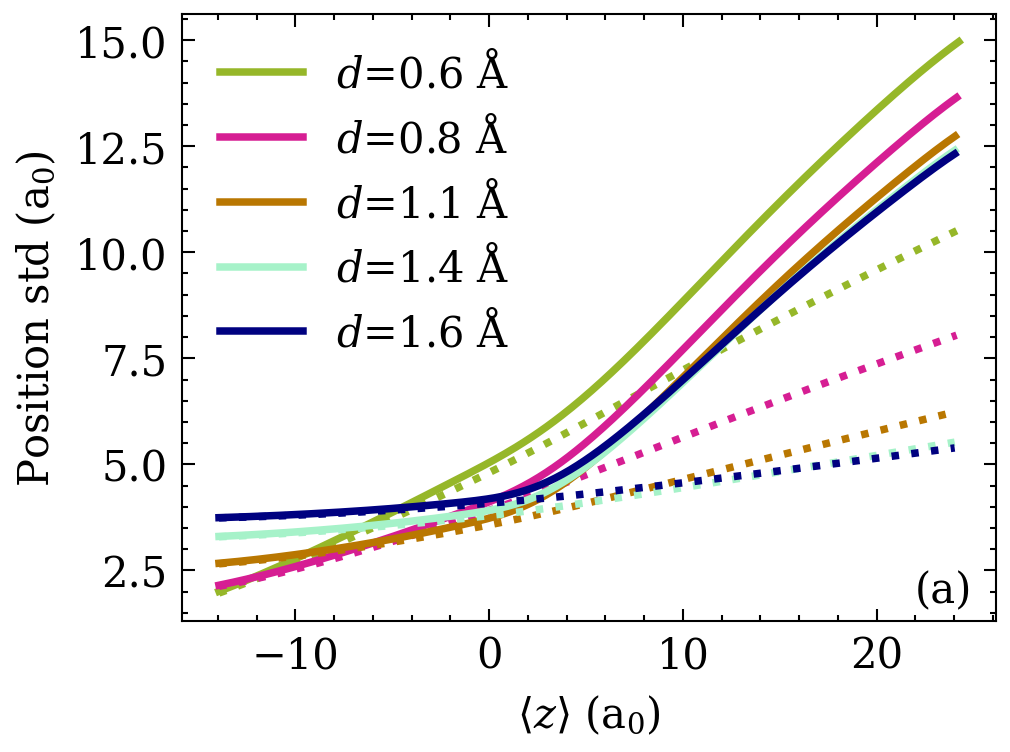}
    \includegraphics[width=0.46\linewidth]{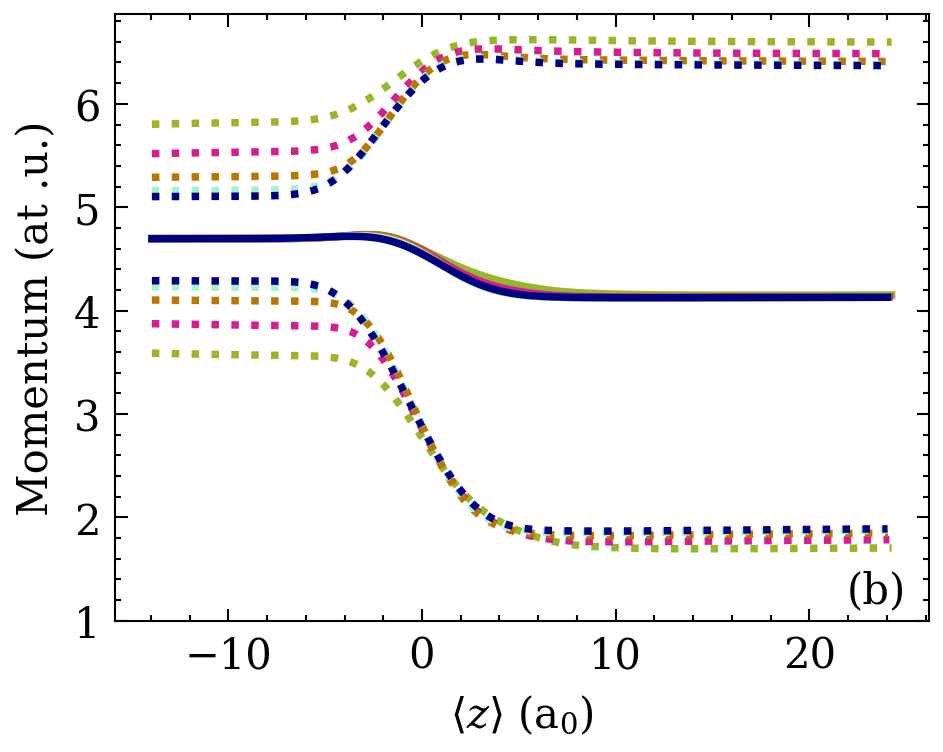}
    \caption{(a) The evolution of the real space distribution of the wave packets with different initial widths, before and after impacting graphene at ($\left<z\right>=0\, a_0$). The solid lines are the position standard deviation calculated by rt-TDDFT, and the dotted lines are the analytical solution in the vacuum, as denoted in Eq.~\ref{eq:vac wp std}.
    (b) The momentum space distribution for wave packets with different widths. The solid line is the mean momentum, and the dotted lines are the mean plus standard deviation of the momentum. }
    \label{fig:wp pos vel std com}
\end{figure}

In the main text, all the electronic wave packets are initilialized with a fixed width of $d=1.1\,\mathrm{\AA}$. In this section, we further investigate 
how varying the initial width affects the evolution of the wave packet in both real and momentum space.

Prior to interacting with the target, the real-space standard deviation of each wave packet follows their analytical solution in the vacuum given in Eq.~\ref{eq:vac wp std}. 
After the impact, the real-space standard deviations of the wave packets tend to align, evolving roughly in parallel. 
This observation is complemented in momentum space, where the mean and standard deviation also converge post-impact, agreeing with the real-space distribution. Despite this convergence in momentum space, the real-space distributions retain some dependence on the initial width and do not become identical.

\begin{figure}
    \centering
    \includegraphics[width=0.45\linewidth]{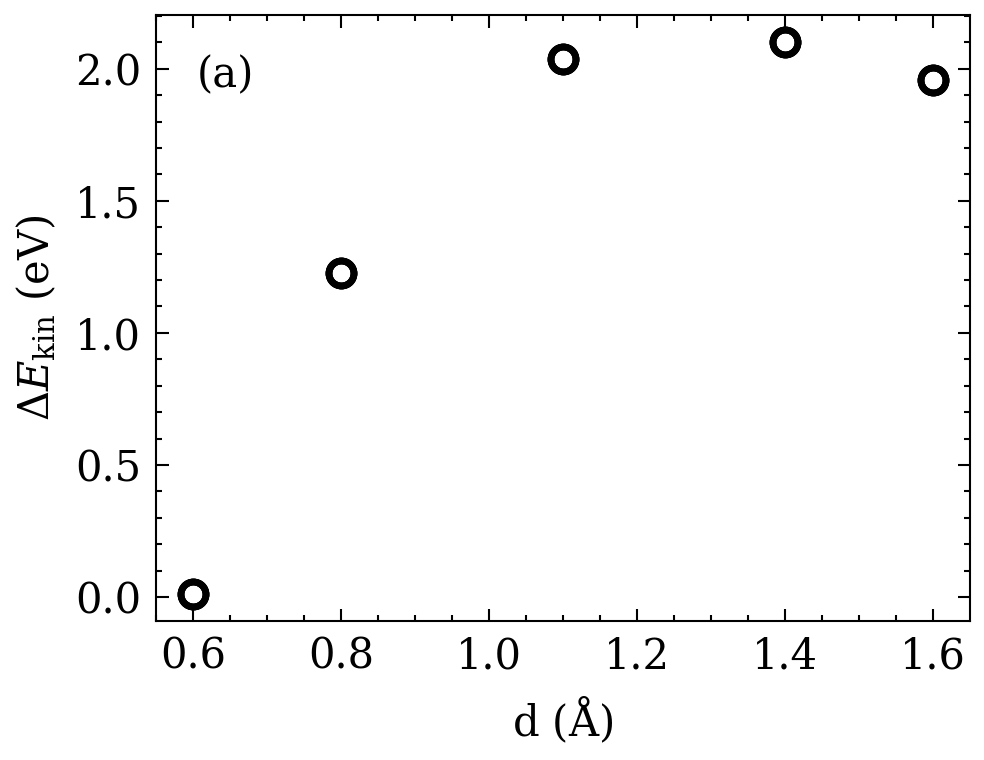}
    \includegraphics[width=0.45\linewidth]{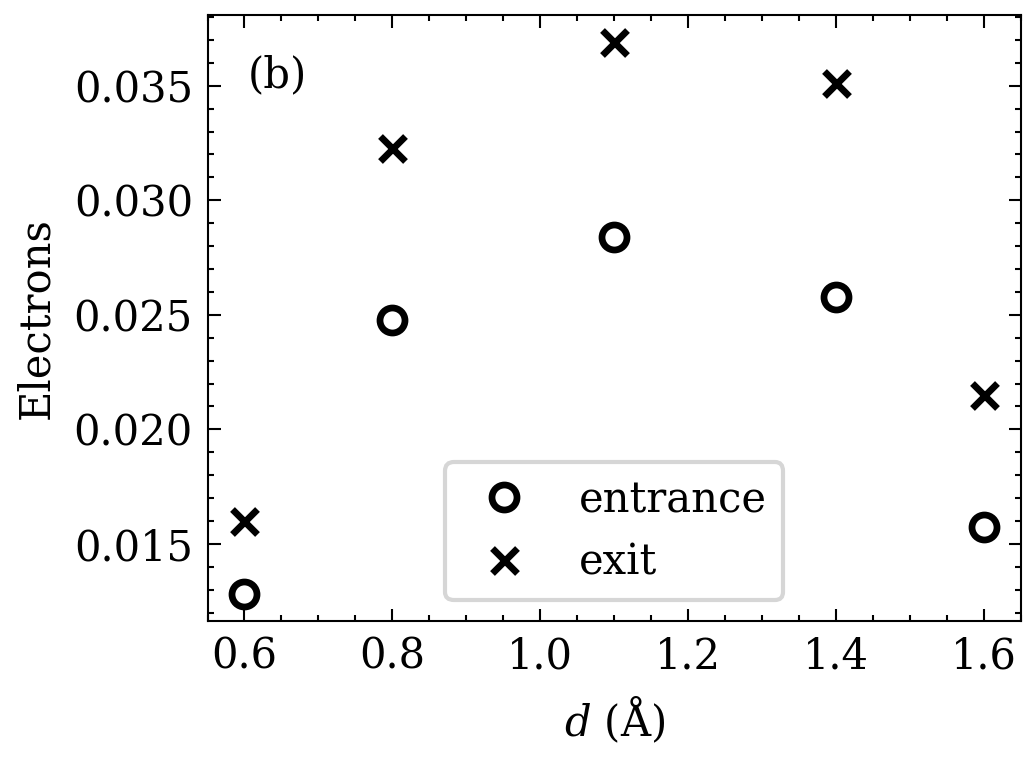}
    \caption{Comparison of (a) the kinetic energy loss and (b) the secondary electrons from both sides of the vacuum with different widths of the wave packet.}
    \label{fig:wp dke sec std com}
\end{figure}

We further explore how the initial width influences secondary electron emission. 
Fig.~\ref{fig:wp dke sec std com} denotes that the kinetic energy loss varies significantly across different initial widths.
This finding aligns with the momentum evolution depicted in Fig.~\ref{fig:wp pos vel std com}\,b: although the final momentum distributions are similar among all wave packets, their distinct initial states lead to different amounts of kinetic energy loss.
Such differences directly impact secondary electron emission; for instance, the secondary electron yield for $d = 1.1\,\mathrm{\AA}$ can be nearly twice that for $d = 0.6\,\mathrm{\AA}$, as illustrated in Fig.~\ref{fig:wp dke sec std com}. This introduces an element of uncertainty in our simulations that warrants further investigation.

Finally, we note that our target material, graphene, consists of a single atomic layer. For thicker systems—such as multilayer graphene, thin films, or bulk materials, we expect the influence of the initial wave packet state on the interaction dynamics to diminish rapidly. 
This relative insensitivity to the initial state has important implications for simulating electron irradiation in bulk systems, where the incident wave packet must be first orthonormalized against the Kohn–Sham states of the target. Although this orthonormalization inevitably distorts the initial real- and momentum-space distributions of the wave packet, the observed convergence of the electron dynamics despite variations in the initial state suggests that this distortion may have a limited impact on the subsequent electron dynamics, especially after the initial transients have dissipated. 
This electron irradiation in the bulk materials with different initial state could be of interest in the future study.

\end{document}